\documentclass[aps,prd,preprint,groupedaddress,nofootinbib]{revtex4-1}

\newcommand{\avg}[1]{\left< #1 \right>} 
\usepackage{graphicx}
\usepackage{caption}
\usepackage{subcaption}

\usepackage{epstopdf}
\usepackage{rotating}
\usepackage[breaklinks=true,colorlinks=true,linkcolor=blue,urlcolor=blue,citecolor=blue]{hyperref}
\usepackage{rotating}
\usepackage{amsmath,amsfonts,amsthm}

\usepackage[utf8]{inputenc}
\usepackage{tikz-feynman}
\tikzfeynmanset{compat=1.1.0}

\newcommand{\ZZ}{\mathbb{Z}}

\begin{document}
\preprint{DESY 23-191}
	
	
	
\title[]{UV-complete Gauged Anomaly-free U(1) Froggatt-Nielsen Model}

\author{Johan Rathsman}
\email[]{johan.rathsman@fysik.lu.se}
\affiliation{Department of Physics, Lund University, SE-223 62, Lund, Sweden}

\author{Felix Tellander}
\email[]{felix@tellander.se}
\affiliation{Deutsches Elektronen-Synchrotron DESY,\\ Notkestr.~85, 22607 Hamburg, Germany.}

	
	
\date{\today}
	
\begin{abstract}
	We investigate the possibility of understanding all fermion masses and mixings within a gauged Froggatt-Nielsen framework. Continuing the work from [J. Rathsman and F. Tellander, Phys. Rev. D \textbf{100}, 055032 (2019)] we especially focus on a UV completion of this type of models. 
	
	Independent of the UV completion, we construct an anomaly-free two Higgs doublet model with a gauged $U(1)_F$ flavor symmetry and three right-handed neutrinos explaining all observed masses and mixings in the fermion sector. We then investigate two different UV completions: one through fermions and one through scalars. The fermion completion has low lying Landau poles in the gague couplings while the scalar completion is viable up to the gravity scale.

\end{abstract}
	
	
\maketitle
	
\section{Introduction}
\label{intro}
Understanding the flavor structure of the Standard Model (SM) is one of the big challenges of modern particle physics and constitutes one of the best windows into new physics. In this paper we use an anomaly-free gauged $U(1)_F$ flavor symmetry to generate the observed fermion masses and mixings via the Froggatt-Nielsen (FN) mechanism \cite{Froggatt1979}. We implement this mechanism in the minimal way by assuming that all fermions (neutrinos included) have Dirac masses. Thus all hierarchies and mixings are generated by FN suppression of Yukawa couplings. We prove, however, that with only the SM particle content at the electroweak scale, this setup is inconsistent and that no rational $U(1)_F$ charge assignment exists (both with and without Dirac neutrinos). The simplest solution to this problem is to extend the Higgs sector to a two Higgs-doublets model (2HDM) and here we find a model reproducing all observed masses and mixings. Natural flavor conservation is implemented in the Higgs sector from a type-II (MSSM-like) $\ZZ_2$ symmetry which is derived from the $U(1)_F$ symmetry. 

In practice, the FN mechanism is often used in a way such that the physics at the scale $\Lambda_{FN}$ is not specified. When the flavor group is spontaneously broken by a scalar (the flavon $S$) obtaining a vacuum expectation value, a small order parameter $\epsilon=\avg{S}/\Lambda_{FN}$ will be introduced in the low-energy operators. The number of $\epsilon$-suppressions will determine the masses and mixings of the low-energy fermions. However, using vector-like fermions as the UV-completion, as originally proposed in \cite{Froggatt1979}, leads to Landau poles below the Planck scale for the gauge couplings. This was studied in the case of  supersymmetry in \cite{LEURER1993,LEURER1994}. They deduced that the $\Lambda_{FN}$ scale could be of the order of some TeVs, but for most fermion completions much higher. 

In our case with a gauged $U(1)_F$ flavor gauge group, the associated gauge boson will contribute to rare flavor changing processes putting stringent bounds on its mass and coupling. This in turn puts bounds on the scale $\Lambda_{FN}$ by the approximate gauge boson mass relation
\begin{equation}
M_F\approx g_F\avg{S}=g_F\epsilon\Lambda_{FN}
\end{equation}
where $M_F$ and $g_F$ are the mass and coupling of the new boson. The mass (coupling) of the boson may always be chosen large (small) enough so it decouples from experimental observables. But this is an uninteresting scenario so we will keep $\Lambda_{FN}\approx 10^5$ GeV in order to keep the door open for possible observable effects in flavor physics. More detailed phenomenological studies of the Froggatt-Nielsen model and the relation to flavor anomalies have recently been considered in e.g. \cite{Asadi:2023ucx,Cornella:2023zme}.

Another possibility of UV-completion is through addition scalars as done by Bijnens and Wetterich in \cite{Bijnens1987}. This affects the running of the gauge couplings less but introduces the risk of a strongly coupled scalar sector.

Using the implementation of real triangularization in the \texttt{RegularChains} Maple package \cite{RegularChains} (cf. \cite[\S4]{chen2011semi}, \cite{chenThesis} and \cite{chen2007comprehensive})  we obtain an anomaly-free rational $F$-charge assignment reproducing the flavor phenomenology (fermion masses and mixings) via the FN mechanism. We note, that recently methods to this end that do not rely on cylindrical algebraic decomposition (CAD) has been proposed \cite{helmer2023effective}. A Bijnens-Wetterich scalar completion allows $\Lambda_{FN}$ small without Landau poles in the gauge sector.

The rest of this paper is organized as follows. In Section \ref{sec: gauged FN} we described the gauged Froggatt-Nielsen model and the constraints from vanishing gauge anomalies. We show how these are combined into sum rules that the $F$-charges must satisfy. In Section \ref{section: standard model} we consider only the Standard Model as the low-energy theory and conclude that an extended model is needed. This extension is the two-Higgs-doublet model described in Section \ref{saection 2HDM}. The different UV-completions are studied in Section \ref{section: UV} and we conclude the paper in Section \ref{section: discussion}.
\section{Gauged Froggatt-Nielsen model}\label{sec: gauged FN}
The extended SM with three right-handed neutrinos is described by the Yukawa Lagrangian
\begin{equation}
-\mathcal{L}_Y=\overline{Q}_L\widetilde{\Phi}Y^UU_R+\overline{Q}_L\Phi Y^DD_R+\overline{L}_L\Phi Y^LE_R+\overline{L}_L\widetilde{\Phi}Y^NN_R + \mathrm{H.c}\\
\end{equation}
where the family index is suppressed. In the Froggatt-Nielsen framework the mass matrices; $M^F\ \mathrm{with}\ F=U,D,L,N$, are given by \cite{Froggatt1979}
\begin{equation}
M_{ij}^F=\avg{\Phi}_0Y_{ij}^F=\avg{\Phi}_0g_{ij}^F\left(\frac{\avg{S}}{\Lambda_{FN}}\right)^{|n_{ij}^F|}
\end{equation}
where $\avg{S}$ is the vacuum expectation value of the flavon breaking the $U(1)_F$ symmetry, $|n_{ij}^F|$ is the number of $S/S^\dagger$-insertions needed for $F$-invariance and $g_{ij}^F$ are random complex numbers of order one. We define $\avg{S}/\Lambda_{FN}=\epsilon$ and assume $\epsilon\approx 0.2$ to fit with the Wolfenstein parameterization of the CKM matrix \cite{Wolfenstein1983}.

Let the $F$-charges of the left-handed fermion fields $\{Q_L^i,(U_R^i)^c,(D_R^i)^c,L_L^i,(E_R^i)^c,(N_R^i)^c\}$ and the Higgs doublet $\Phi$ be  denoted by $\{Q_i,u_i,d_i,L_i,e_i,\nu_i\}$ and $H$ respectively. The cancellation of gauge anomalies gives the following constraints \cite[Chapter 22.4]{Weinberg1996}
\begin{equation}\label{eq: anomalies U1}
\begin{array}{ll}
\mathcal{A}_{1FF}&=2 {\displaystyle \sum_{j=1}^3}\left(Q_j^{2}-2u_j^{2}+d_j^{2}-L_j^{2}+e_j^{2}\right)=0\\
\mathcal{A}_{11F}&=\dfrac{2}{3} {\displaystyle\sum_{j=1}^{3}}\left(Q_j+8u_j+2d_j+3L_j+6e_j\right)=0\\
\mathcal{A}_{33F}&=\dfrac{1}{2}{\displaystyle\sum_{j=1}^3}\left(2Q_j+u_j+d_j\right)=0\\
\mathcal{A}_{22F}&=\dfrac{1}{2}{\displaystyle\sum_{j=1}^3}\left(3Q_j+L_j\right)=0\\
\mathcal{A}_{FFF}&= {\displaystyle\sum_{j=1}^3}\left(6Q_j^{3}+3u_j^{3}+3d_j^{3}+2L_j^{3}+e_j^{3}+\nu_j^3\right)=0\\
\mathcal{A}_{ggF}&={\displaystyle\sum_{j=1}^3}\left(6Q_j+3u_j+3d_j+2L_j+e_j+\nu_j\right)=0
\end{array}
\end{equation}
where $\mathcal{A}_{ggF}$ is from the triangle diagram with two gravitons and one $U(1)_F$ gauge boson. Moreover, the FN constraints for $F-$invariant Yukawa couplings are given by
\begin{equation}\label{eq: FN constraints}
\begin{array}{ll}
n_{ij}^U=Q_i+u_j+H,\ \ \ &n_{ij}^N=L_i+\nu_j+H\\
n_{ij}^D=Q_i+d_j-H,\ \ \ &n_{ij}^L=L_i+e_j-H
\end{array}
\end{equation}
where number of flavon insertions is given by $|n_{ij}^F|$. To a leading order approximation the Yukawa matrices are diagonalized by the biunitary transformations
\begin{align}
Y^U&=(V_L^U)^\dagger D^UV_R^U,\ \ \ Y^D=(V_L^D)^\dagger D^DV_R^D,\nonumber\\
Y^L&=(V_L^L)^\dagger D^LV_R^L,\ \ \ Y^N=(V_L^N)^\dagger D^NV_R^N,\nonumber
\end{align}
where $D^{F},\ F=U,D,L,N$ are diagonal and the left matrices are given by
\begin{align}
(V_L^{U,D})_{ij}&\sim \epsilon^{|Q_i-Q_j|},\ \ \ (V_L^{L,N})_{ij}\sim \epsilon^{|L_i-L_j|}
\end{align}
while the right matrices are
\begin{align}
(V_R^{U})_{ij}\sim \epsilon^{|u_i-u_j|},\ \ \ (V_R^{D})_{ij}&\sim \epsilon^{|d_i-d_j|},\ \ \ (V_R^{L})_{ij}\sim \epsilon^{|e_i-e_j|},\ \ \ (V_R^{N})_{ij}\sim \epsilon^{|\nu_i-\nu_j|}.
\end{align}
This gives the mixing matrices
\begin{equation}
V_{CKM}=V_L^U(V_L^D)^\dagger\sim\epsilon^{|Q_i-Q_j|},\ \ \ U_{PMNS}=V_L^N(V_L^L)^\dagger\sim\epsilon^{|L_i-L_j|}.
\end{equation}

As was shown in \cite{Tellander2019} there is a non-trivial relation between the FN constraints and the anomaly constraints, which manifest as sum rules for the $n_{ij}^F$. For the SM with three right-handed neutrinos there are three independent sum rules:
\begin{equation}\label{eq: sum rules}
\begin{array}{l}
{\displaystyle \sum_{i=1}^3}(n_{ii}^U+n_{ii}^D)=2\mathcal{A}_{33F}=0\\
{\displaystyle  \sum_{i=1}^3}(n_{ii}^D-n_{ii}^L)=\frac{8}{3}\mathcal{A}_{33F}-\frac{1}{4}\mathcal{A}_{11F}-\mathcal{A}_{22F}=0\\
{\displaystyle \sum_{i=1}^3}(n_{ii}^L+n_{ii}^N)=\mathcal{A}_{ggF}-6\mathcal{A}_{33F}=0
\end{array}
\end{equation}
which must be satisfied. 
\section{Standard Model}\label{section: standard model}
Running the fermion masses in the SM to 100 TeV, which is our assumed scale $\Lambda_{FN}$ of new physics, gives 
\begin{equation}
\begin{array}{lll}
m_u=8.9\cdot 10^{-4}\ \mathrm{GeV},&\qquad m_c=0.41\ \mathrm{GeV},&\qquad m_t=120\ \mathrm{GeV},\\
m_d=2.0\cdot 10^{-3}\ \mathrm{GeV},&\qquad m_s=0.037\ \mathrm{GeV},&\qquad m_b=1.9\ \mathrm{GeV},\\
m_e=4.8\cdot 10^{-4}\ \mathrm{GeV},&\qquad m_\mu=0.1\ \mathrm{GeV},&\qquad m_\tau=1.7\ \mathrm{GeV}
\end{array}
\end{equation} 
using the starting values from \texttt{2HDME} \cite{Oredsson:2018vio}. This corresponds to the powers
\begin{equation}
\begin{array}{ccc}
|n_{11}^U|=8,&\qquad |n_{22}^U|=4,&\qquad|n_{33}^U|=0,\\
|n_{11}^D|=7,&\qquad |n_{22}^D|=5,&\qquad|n_{33}^D|=3,\\
|n_{11}^L|=8,&\qquad |n_{22}^L|=5,&\qquad|n_{33}^L|=3.
\end{array}
\end{equation} 
Clearly none of the sum rules may be satisfied so we conclude that the SM field content is not enough. We remark that one way to satisfy the sum rules is to chose $|n_{22}^L|=4$, thus $g_{22}^L\approx 0.378$ and $|n_{11}^U|=19$. However, it is still impossible to find rational flavon charges, which can be proved by real triangularization.

Adding Dirac neutrinos to the SM will not affect the fact that $|n_{11}^U|$ has to be 19 for the sum rules to be satisfied, but it introduces more relations among the equations and by doing so makes $\mathcal{A}_{FFF}$ redundant, thus we may find rational solutions. We will work in a three neutrino paradigm with normal hierarchy: $m_{\nu_1} \ll m_{\nu_2}<m_{\nu_3}$, yielding $m_{\nu_3}\approx 0.0506$ eV, $m_{\nu_2}\approx 0.0086$ eV and $m_{\nu_1}$ arbitrary as long as it is much smaller than the other two. This corresponds to $\epsilon^{18}$ and $\epsilon^{19}$ for the two heaviest neutrinos. Moreover, we assume $L_2-L_3=0$ for large $\nu_\mu-\nu_\tau$ mixing. This implies that the only possible mixing with the first generation is given by $|L_1-L_2|=115$, which is practically insignificant and phenomenologically unfeasible. 

\section{ Two Higgs Doublet Model}\label{saection 2HDM}
To resolve the issues with implementing the FN-mechanism in the SM we add an additional Higgs doublet. The reason this works is that for a 2HDM there are only two independent sum rules (one if we do not have Dirac neutrinos).

It was shown in \cite[Fig. 7]{Oredsson2019} that for a 2HDM with exact $\mathbb{Z}_2$ symmetry to be stable up to (at least) 100 TeV it is necessary for $\tan\beta\approx 1$. There are different choices of $\mathbb{Z}_2$ symmetry, each leading to different sum rules, however, the only choice we found that works is type-II (MSSM).

 Running the Yukawa couplings to 100 TeV we get (see e.g. \cite{Plantey2019})
\begin{equation}
\begin{array}{ccc}
|n_{11}^U|=7,&\qquad |n_{22}^U|=3,&\qquad|n_{33}^U|=0,\\
|n_{11}^D|=7,&\qquad |n_{22}^D|=5,&\qquad|n_{33}^D|=3,\\
|n_{11}^L|=8,&\qquad |n_{22}^L|=4,&\qquad|n_{33}^L|=3.
\end{array}
\end{equation} 
However to reproduce the neutrino sector, specifically $L_1-L_2=0$, we have to change $|n_{11}^U|$ to 8 (corresponding to $g_{11}^U\approx2.94$), $|n_{33}^D|=2$ (corresponding to $g_{33}^D\approx 0.39$) and $|n_{33}^L|=2$ (corresponding to $g_{33}^L\approx 0.35$). All these are minor changes and well within the $g_{ij}^F\sim\mathcal{O}(1)$ paradigm.

The textures we want to reproduce are
\begin{equation}\label{eq: Yukawa}
Y^U\sim\begin{pmatrix}
\epsilon^8 & \epsilon^5 & \epsilon^3\\
\epsilon^7 & \epsilon^4 & \epsilon^2\\
\epsilon^5 & \epsilon^2 & 1
\end{pmatrix},\ Y^D\sim\begin{pmatrix}
\epsilon^7 & \epsilon^6 & \epsilon^5\\
\epsilon^6 & \epsilon^5 & \epsilon^4\\
\epsilon^4 & \epsilon^3 & \epsilon^2
\end{pmatrix},\ Y^L\sim\begin{pmatrix}
\epsilon^8 & \epsilon^4 & \epsilon^2\\
\epsilon^8 & \epsilon^4 & \epsilon^2\\
\epsilon^8 & \epsilon^4 & \epsilon^2
\end{pmatrix},\ Y^N\sim\begin{pmatrix}
\epsilon^{|a|} & \epsilon^{19} & \epsilon^{18}\\
\epsilon^{|a|} & \epsilon^{19} & \epsilon^{18}\\
\epsilon^{|a|} & \epsilon^{19} & \epsilon^{18}
\end{pmatrix}.
\end{equation}
with mixing matrices
\begin{equation}
V_{CKM}\sim\begin{pmatrix}
1 & \epsilon & \epsilon^3\\
\epsilon & 1 & \epsilon^2\\
\epsilon^3 & \epsilon^2 & 1
\end{pmatrix},\qquad U_{PMNS}\sim\begin{pmatrix}
1&1&1\\
1&1&1\\
1&1&1
\end{pmatrix}
\end{equation}
where $a$ is an integer determining the mass of the lightest neutrino (to fit with normal hierarchy we need $|a|>19$). We implement this as solving the semi-algebraic system
\begin{align*}
(Q_1-Q_2)^2 &= 1,\qquad (Q_2-Q_3)^2 = 4,\qquad (Q_1-Q_3)^2 = 9,\\
(Q_1+u_1+H_2)^2 &= 8^2,\qquad (Q_2+u_2+H_2)^2 = 4^2,\qquad Q_3+u_3+H_2 = 0,\\
(Q_1+d_1-H_1)^2 &= 7^2,\qquad (Q_2+d_2-H_1)^2 = 5^2,\qquad (Q_3+d_3-H_1)^2 = 2^2,\\
(L_1+e_1-H_1)^2 &= 8^2,\qquad (L_2+e_2-H_1)^2 = 4^2,\qquad (L_3+e_3-H_1)^2 = 2^2,\\
(L_1+\nu_1+H_2)^2 &= a^2,\qquad (L_2+\nu_2+H_2)^2 = 19^2,\qquad (L_3+\nu_3+H_2)^2 = 18^2,\\
L_2-L_3 &= 0,\qquad L_1-L_2 = 0,\\
\mathcal{A}_{1FF}=0,\ \ \mathcal{A}_{11F}&=0,\ \ \mathcal{A}_{33F}=0,\ \ \mathcal{A}_{22F}=0,\ \ \mathcal{A}_{FFF}=0,\ \ \mathcal{A}_{ggF}=0,\\
(Q_1+u_3+H_2)^2 &\ge (Q_2+u_3+H_2)^2,\qquad (Q_2+u_3+H_2)^2 \ge (Q_3+u_3+H_2)^2,\\
(Q_1+u_2+H_2)^2 &\ge (Q_2+u_2+H_2)^2,\qquad (Q_2+u_2+H_2)^2 \ge (Q_3+u_2+H_2)^2,\\
(Q_1+d_3-H_1)^2 &\ge (Q_2+d_3-H_1)^2,\qquad (Q_2+d_3-H_1)^2 \ge (Q_3+d_3-H_1)^2,\\
(Q_1+d_2-H_1)^2 &\ge (Q_2+d_2-H_1)^2,\qquad (Q_2+d_2-H_1)^2 \ge (Q_3+d_2-H_1)^2
\end{align*}
which we solve by real triangularization in the \texttt{RegularChains} Maple package \cite{RegularChains}. This yields two values of $|a|$: 25 and 49. Meaning that this model predicts the lightest neutrino mass to
\begin{equation}
m_{\nu_1}\sim\epsilon^{25}\cdot174\ \mathrm{GeV}\approx 10^{-6}\ \mathrm{eV}
\end{equation}
or
\begin{equation}
m_{\nu_1}\sim\epsilon^{49}\cdot174\ \mathrm{GeV}\approx 10^{-23}\ \mathrm{eV}.
\end{equation}
One of the solutions for $|a|=25$ is given by
\[\arraycolsep=1.1pt\def\arraystretch{1.4}
\begin{array}{lll}
Q_1=\dfrac{82111}{51651}+\dfrac{1}{3}H_2,\qquad &Q_2=\dfrac{30460}{51651}+\dfrac{1}{3}H_2,\qquad &Q_3=-\dfrac{72842}{51651}+\dfrac{1}{3}H_2,\\
u_1=\dfrac{331097}{51651}-\dfrac{4}{3}H_2,\qquad&u_2=\dfrac{176144}{51651}-\dfrac{4}{3}H_2,\qquad&u_3=\dfrac{72842}{51651}-\dfrac{4}{3}H_2,\\
d_1=-\dfrac{168196}{51651}+\dfrac{2}{3}H_2,\qquad& d_2=-\dfrac{219847}{51651}+\dfrac{2}{3}H_2, & d_3=-\dfrac{271498}{51651}+\dfrac{2}{3}H_2,\\
L_1=L_3,\qquad& L_2=L_3, \qquad& L_3=-\dfrac{13243}{17217}-H_2,\\
e_1=\dfrac{1765}{17217}+2H_2,\qquad & e_2=-\dfrac{67103}{17217}+2H_2,\qquad & e_3=-\dfrac{101537}{17217}+2H_2,\\
\nu_1=-\dfrac{417182}{17217},\qquad & \nu_2=\dfrac{340366}{17217},\qquad & \nu_3=\dfrac{323149}{17217},\\
H_1=H_2-\dfrac{26}{3},\qquad & H_2\in\mathbb{Q}
\end{array}
\]
where we note that $H_2$ is a free rational number. This number may be fixed by adding some extra constraint, e.g. removing mixing between $U(1)_Y$ and $U(1)_F$ in the massless limit by adding
\begin{equation}\label{eq: kinetic mixing}
\sum_{j=1}^3 (2Q_j-4u_j+2d_j-2L_j+2e_j)=0
\end{equation}
to the list. This is just the trace of the hyper charge and $F$-charge generators. The $F$-charges are now uniquely specified and given by
\begin{equation}\label{eq: charges}
\begin{array}{lll}
Q_1=196891/86085,&\qquad Q_2=110806/86085,&\qquad Q_3=-61364/86085,\\
u_1=311671/86085,&\qquad u_2=53416/86085,&\qquad u_3=-118754/86085,\\
d_1=-53416/28695,&\qquad d_2=-82111/28695,&\qquad d_3=-110806/28695,\\
L_1=-82111/28695,&\qquad L_2=-82111/28695,&\qquad L_3=-82111/28695,\\
e_1=369061/86085,&\qquad e_2=24721/86085,&\qquad e_3=-147449/86085,\\
\nu_1=-417182/17217,&\qquad \nu_2=340366/17217,&\qquad \nu_3=323149/17217,\\
H_1=-565952/86085,&\qquad H_2=180118/86085.
\end{array}
\end{equation}
Whenever specific charges are used in the following, it will be these charges we use.
\section{UV-completion}\label{section: UV}
\begin{figure*}%
\captionsetup[subfigure]{justification=centering}
\newcommand{\xs}{.8}%
    \begin{subfigure}[b]{0.5\textwidth}
        \centering
         \begin{tikzpicture}[baseline=-\the\dimexpr\fontdimen22\textfont2\relax]
		\begin{feynman}[scale=1.1,transform shape]
		\vertex (a1);
		\vertex [above left = of a1] (b1);
		\vertex [below left = of a1] (c1);
		\vertex [right = of a1] (a2);
		\vertex [above = of a2] (b2);
		\vertex [right = of a2] (a3);
		\vertex [below = of a3] (d1);
		\vertex [right = of a3] (a4);
		
		\diagram* {
			(b1) -- [fermion, edge label=\(Q_{L}\)] (a1),
			(c1) -- [charged scalar, insertion=0, edge label=\(\Phi\)] (a1),
			(a1) -- [insertion=0.5, edge label=\(\Lambda_{FN}\)] (a2),
			(b2) -- [charged scalar, insertion=0, edge label=\(S\)] (a2),
			(a2) -- [insertion=0.5, edge label=\(\Lambda_{FN}\)] (a3),
			(a3) -- [fermion, edge label=\(U_R\)] (a4),
			(d1) -- [charged scalar, insertion=0, edge label=\(S\)] (a3),
		};
		\end{feynman}
		\end{tikzpicture}
        \caption{}
        \label{fig: fermion}
    \end{subfigure}\hfill
    \begin{subfigure}[b]{0.5\textwidth}
        \centering
         \begin{tikzpicture}[baseline=-\the\dimexpr\fontdimen22\textfont2\relax]
		\begin{feynman}[scale=1.1,transform shape]
		\vertex (a1);
		\vertex [above left = of a1] (b1);
		\vertex [below left = of a1] (c1);
		\vertex [right = of a1] (a2);
		\vertex [above = of a2] (b2);
		\vertex [right = of a2] (a3);
		\vertex [below = of a3] (c2);
		\vertex [right = of a3] (a4);
		
		\diagram* {
			(b1) -- [fermion, edge label=\(Q_{L}\)] (a1),
			(c1) -- [fermion, edge label=\((U_R)^c\)] (a1),
			(a1) -- [charged scalar, edge label=\(\widetilde{\Phi}_2\)] (a2),
			(a2) -- [charged scalar, edge label=\(\widetilde{\Phi}_1\)] (a3),
			(a3) -- [charged scalar, insertion=1, edge label=\(\widetilde{\Phi}\)] (a4),
			(b2) -- [charged scalar, insertion=0, edge label=\(S\)] (a2),
			(c2) -- [charged scalar, insertion=0, edge label=\(S\)] (a3),
		};
		\end{feynman}
		\end{tikzpicture}
        \caption{}
        \label{fig: scalar}
    \end{subfigure}
   \caption{Generating an $\epsilon^2$ suppressed mass using the FN mechanism with vector-like fermions (\subref{fig: fermion}) and using a Bijnens-Wetterich scalar completion (\subref{fig: scalar}).}
	\label{figure: top mass}
\end{figure*}
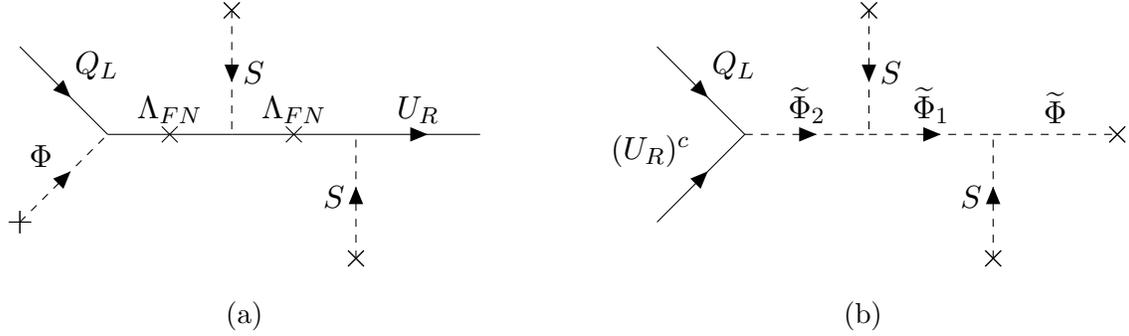

In their original work, Froggatt and Nielsen suggested that the physics at the $\Lambda_{FN}$ scale could be vector-like fermions. Later and independently, Bijnens and Wetterich suggested that the same type of hierarchy mechanism could be achieved from a scalar UV-completion \cite{Bijnens1987}. One way of generating an $\epsilon^2$ suppressed mass for each case is shown in Fig. \ref{figure: top mass}. We note that the results and conclusions in this section are independent of $U(1)_F$ being gauged or not.

The fermion completion was studied in a supersymmetric case in \cite{LEURER1993,LEURER1993} and it was found that one in general gets Landau poles below the Planck scale if $\Lambda_{FN}$ is not chosen large enough (sometimes as large as $10^8$ TeV but other times as small as 100 TeV). We use the following convention for the renormalization group equation (RGE)
\begin{equation}
\mu\frac{d}{d\mu}\alpha=-\frac{\alpha^2}{2\pi}\beta
\end{equation}
where $\alpha=g^2/4\pi$ and the one-loop beta function is given by
\begin{equation}
\beta_G=\frac{11}{3}C_2(G)-\frac{2}{3}n_WT(R_W)-\frac{1}{3}n_sT(R_s)
\end{equation}
where $C_2(G)=N$ for $G=SU(N)$ ($C_2(G)=0$ if $G=U(1)$), $n_W$ is the number of Weyl fermions in representation $R_W$ where $T(R_W)$ satisfies $\mathrm{tr}(T_W^aT_W^b)=T(R_W)\delta^{ab}$ and $n_s$ is the number of complex scalars in representation $R_s$.

\begin{figure}
	\centering
	\includegraphics[width=0.99\textwidth]{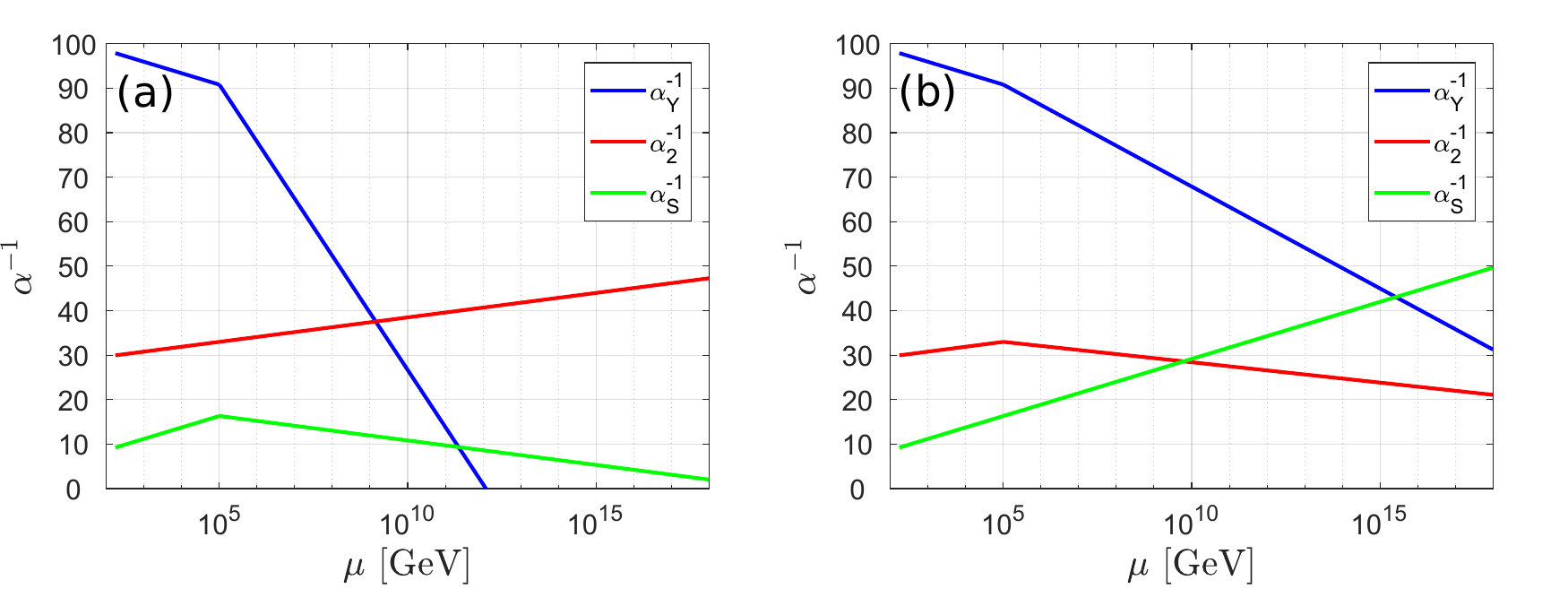}
	\caption{(a) One-loop evolution of the SM gauge couplings with fermion completion at $\Lambda_{FN}= 100$ TeV. (b) One-loop evolution of the SM gauge couplings with scalar completion at $\Lambda_{FN}=100$ TeV. }
	\label{fig: RGE}
\end{figure}

Below $\Lambda_{FN}$ we have a 2HDM with beta functions
\begin{equation}
\beta_Y=-7,\ \ \ \beta_2=3,\ \ \ \beta_S=7.
\end{equation}
The Yukawa couplings in Eq. (\ref{eq: Yukawa}) tell us that the fermion content above $\Lambda_{FN}$ has to support 8 flavon insertions in the up sector, 7 insertions in the down etc. In total this amounts to adding (counting the Weyl fermions); 8 new up-type quarks, 7 down-type quarks (each quark with three possible color states) and 8 leptons.\footnote{The fermions needed for the neutrino masses do not affect the running since they are SM singlets.}
The necessary vector-like fermion content  changes the one-loop beta-functions to 
\begin{equation}\label{eq: beta fnc fermion}
\beta_Y=-35,\ \ \ \beta_2=3,\ \ \ \beta_S=-3
\end{equation}
above $\Lambda_{FN}=100$ TeV. The RGE evolution for this UV-completion is shown in Fig. \ref{fig: RGE}(a), which has a Landau pole for $g_Y$ at $10^{12}$ GeV. This could be avoided by pushing $\Lambda_{FN}$ higher but then the $U(1)_F$ boson and the FN fermions decouple so that they are not observable in any experiment. For us this is not a satisfying solution.

Switching to a scalar Bijnens-Wetterich completion, we need one chain of (including the 2HDM doublets) 9 doublets to reproduce the down-type quarks and charge lepton masses and 26 doublets to reproduce the up-type quark and neutrino masses. These doublets change the beta functions to 
\begin{equation}
\beta_Y=-\frac{25}{2},\ \ \ \beta_2=-\frac{5}{2},\ \ \ \beta_S=7.
\end{equation}
With this UV-completion there are no Landau poles below the Planck scale as can be seen in Fig. \ref{fig: RGE}(b). It is important to note here that the charges in Eq. (\ref{eq: charges}) have nothing to do with the above argument, the beta-functions are completely determined by the Yukawa couplings in Eq. (\ref{eq: Yukawa}) and the constraints for family mixing (i.e. the structure of the CKM and PMNS matrix).

We have not included the running for $g_F$ since, even in the phenomenologically interesting region, it is possible to chose it small enough to guarantee no Landau pole below the Planck scale; $g_F\lesssim 3\cdot10^{-2}$ works for both completions.   

In view of the running of gauge couplings the scalar completion is clearly the preferred UV-completion. However, multiple scalar doublets generically leads to Landau poles for the quartic couplings (see e.g. \cite[Fig. 3]{Oredsson2019} for the case with two Higgs doublets). With a total of 35 doublets our situation is much more severe than in \cite{Oredsson2019}. 

Above $\Lambda_{FN}$ the scalar potential is given by (see also \cite{Bijnens1987}):
\begin{align}
V(\Phi_{1,1},&\ldots,\Phi_{1,9},\Phi_{2,1},\ldots,\Phi_{2,26},S)=\\
&\sum_{i=1}^9m_{1,i}^2\Phi_{1,i}^\dagger\Phi_{1,i}+\sum_{i=1}^{26}m_{2,i}^2\Phi_{2,i}^\dagger\Phi_{2,i}+m_S^2S^\dagger S+\nonumber\\
&\sum_{i,j,k,l}^9\lambda_{1,ijkl}\left(\Phi_{1,i}^\dagger\Phi_{1,j}\Phi_{1,k}^\dagger\Phi_{1,l}\right)+\sum_{i,j,k,l}^{26}\lambda_{2,ijkl}\left(\Phi_{2,i}^\dagger\Phi_{2,j}\Phi_{2,k}^\dagger\Phi_{2,l}\right)+\lambda_S\left(S^\dagger S\right)^2+\nonumber\\
&\sum_{i=1}^9\lambda_{1S,i}\Phi_{1,i}^\dagger\Phi_{1,i}S^\dagger S+\sum_{i=1}^{26}\lambda_{2S,i}\Phi_{2,i}^\dagger\Phi_{2,i}S^\dagger S+\nonumber\\
&\sum_{i=1}^8\alpha_{1,i}\Lambda_{FN}\left(\Phi_{1,i}^\dagger\Phi_{1,i+1}S+\Phi_{1,i+1}^\dagger\Phi_{1,i}S^\dagger\right)+\sum_{i=1}^{25}\alpha_{2,i}\Lambda_{FN}\left(\Phi_{2,i}^\dagger\Phi_{2,i+1}S+\Phi_{2,i+1}^\dagger\Phi_{2,i}S^\dagger\right)+\nonumber\\
&\sum_{i=1}^7\beta_{1,i}\left(\Phi_{1,i}^\dagger\Phi_{1,i+2}SS+\Phi_{1,i+2}^\dagger\Phi_{1,i}S^\dagger S^\dagger\right)+\sum_{i=1}^{24}\beta_{2,i}\left(\Phi_{2,i}^\dagger\Phi_{2,i+2}SS+\Phi_{2,i+2}^\dagger\Phi_{2,i}S^\dagger S^\dagger\right)\nonumber
\end{align}
where $\lambda_{1,ijkl}=\lambda_{1,kjil}=\lambda_{1,jilk}^*$ and $i+k=j+l$, the same holds for $\lambda_{2,ijkl}$. All fields in the sequence $\{\Phi_{1,i}\}$ have $F$-charge $H_1=-565952/86085$ while all fields in $\{\Phi_{2,i}\}$ have $H_2=180118/86085$. This means that the $F$-charge difference between the two sequences is 26/3 for all fields and therefore there are no terms of the form $m_{12}^2\Phi_{1,i}^\dagger\Phi_{2,j}$ or $\lambda(\Phi_{1,i}^\dagger\Phi_{2,j})^2$ in $V$. Therefore, the low-lying 2HDM has a type-II (MSSM-like) $\ZZ_2$-symmetry. This was of course originally an assumption of our model, but from this point of view it can be seen as a derived effect after integrating out the physics at the $\Lambda_{FN}$ scale.
\section{Discussion}\label{section: discussion}
Here we have shown that the Froggatt-Nielsen mechanism with one gauged $U(1)$ group may be used to generate all observed fermion masses and mixings. This construction even provides two predictions for the order of magnitude of the lightest neutrino mass. 

When trying to specify the UV theory a potential problem arises. With a fermion completion the hypercharge coupling obtains a low laying Landau pole, however, with a scalar completion this problem is circumvented.

From this we can conclude that using the FN mechanism with one $U(1)$ to explain the fermion mass hierarchies and mixings is in principle possible. However, the necessary charges are quite ``unnatural" and a much more detailed phenomenological study is needed. With the complexity of the physics above the $\Lambda_{FN}$ scale, this seems like a daunting task.

\section*{Acknowledgments}
We thank Jonas Wittbrodt and Astrid Ordell for many discussions. 
This work is supported in part by the Swedish Research Council, contract number 2016-05996, the European Research Council (ERC) under the European Union's Horizon 2020 research and innovation programme (grant agreement No 668679) and by the Anders Wall Foundation.
\bibliography{Ref}
	
\end{document}